\documentclass[prb,aps,onecolumn,amsmath,amssymb,floatfix,
superscriptaddress]{revtex4}

\usepackage[dvips]{graphics}
\usepackage{color}
\definecolor{dred}{rgb}{0,0,0.6}

\begin{document}

\title{Externally controlled high degree of spin polarization and spin 
inversion in a conducting junction: Two new approaches}

\author{Moumita Patra}

\affiliation{Physics and Applied Mathematics Unit, Indian Statistical
Institute, 203 Barrackpore Trunk Road, Kolkata-700 108, India}

\author{Santanu K. Maiti$^{*,}$}

\affiliation{Physics and Applied Mathematics Unit, Indian Statistical
Institute, 203 Barrackpore Trunk Road, Kolkata-700 108, India}

\begin{abstract}

We propose two new approaches for regulating spin polarization and spin
inversion in a conducting junction within a tight-binding framework based
on wave-guide theory. The system comprises a magnetic quantum ring with 
finite modulation in site potential is coupled to two non-magnetic 
electrodes. Due to close proximity an additional tunneling is established
between the electrodes which regulates electronic transmission significantly. 
At the same time the phase associated with site potential, which can be
tuned externally yields controlled transmission probabilities. Our results
are valid for a wide range of parameter values which demonstrates the 
robustness of our proposition. We strongly believe that the proposed model
can be realized in the laboratory.

\end{abstract}

\maketitle

The way of getting selective spin transmission through a conducting 
junction has always been an interesting topic in the subject of 
spintronics~\cite{ref1,ref2}. The most common route of generating polarized 
spin currents is the use of ferromagnetic electrodes~\cite{ref4,ref5} though 
it has strong limitations due to resistivity mismatch~\cite{ref3}. 
Utilizing a simple quantum dot (QD) driven by radio frequency gate voltages 
one can also get polarized spin current in presence of moderate in-plane 
magnetic field~\cite{extra1,extra2}.

For purposeful design of spintronics devices like spin filters, spin 
transistors, single spin memories, solid state qubits, etc., the 
generation of polarized spin current is not the only requirement, but
its proper regulation is highly significant~\cite{dev1,th1,th2}. Some 
intrinsic properties, for example, spin-orbit (SO) interaction which 
couples electron's spin to the charge degree of freedom provides deeper 
insight~\cite{intrinsic1,intrinsic2,intrinsic3,intrinsic4,intrinsic5} 
for generating polarized spin current. Usually two types of SO 
interactions, namely Rashba~\cite{rashba} and Dresselhaus~\cite{dressel}, 
are encountered in solid state materials, out of which Rashba SO coupling, 
originated from the lacking of structural symmetry, plays the key role 
for selective spin transfer as one can regulate its coupling strength by 
external gate potential~\cite{gate1,gate2}.

For the {\em three-terminal case} where a bridging material is connected 
with three electrodes this approach is highly 
appreciated~\cite{kim,shelykh,peeters,maiti1}. Whereas for the 
{\em two-terminal system} only SO coupling is not capable for producing 
polarized spin currents as it does not break the Kramer's degeneracy 
between $|k\uparrow\rangle$ and $|-k\downarrow\rangle$ 
states~\cite{ballen,sm1}. Thus one has to incorporate magnetic impurities 
or magnetic field to achieve this goal~\cite{hodd} which essentially brings 
the difficulty as confining a strong magnetic field in a nano-scale region 
such as quantum dot or nano-ring is not so trivial.

Few other approaches have also been discussed to achieve higher degree
of spin polarization. For instance, an organic polymer coupled to a quantum 
\begin{figure}[ht]
{\centering \resizebox*{7.5cm}{4.5cm}{\includegraphics{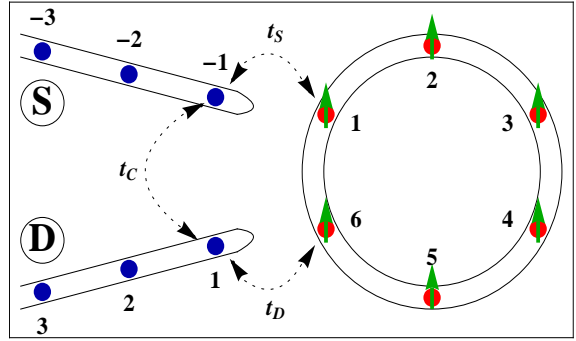}}\par}
\caption{(Color online). Schematic view of conducting nano-junction where 
a magnetic quantum ring with continuous modulation in site energy is 
coupled to two non-magnetic electrodes. Due to close proximity an 
additional {\em new path} is established between source (S) and drain 
(D) electrodes, which is one of the key control parameters of our study. 
Filled colored circles correspond to the atomic sites where 
magnetic atoms, having a finite magnetic moment, are trapped. The direction
of the magnetic moment in each site is described by the green arrow.}
\label{fig1}
\end{figure}
wire can exhibit selective spin transmission~\cite{ext9} where the spin 
polarization is manipulated by an external gate voltage, instead of 
external magnetic field. In another work, Lindelof {\em et al.}, have 
proposed~\cite{ext10} spin reversal in a QD coupled to ferromagnetic leads 
by purely electrical means which provides the fundamental importance of 
designing spintronics devices. Recently one of the authors of us has also 
shown that controlled spin dependent transport can be obtained~\cite{ext11} 
through a magnetic quantum wire coupled to a magnetic quantum ring in 
presence of in-plane electric field. This in-plane electric field regulates 
electronic transport through the junction in a controlled way. 

Till date many works have been done both theoretically as well as
experimentally and have already revealed several unique
features~\cite{ext9,ext10,ext11,texp1,texp2,texp3,texp4,texp5,texp6,texp7,
texp8,texp9,texp10,texp11,texp12,texp13,texp14,texp15} of spin selective 
transmission. 
{\em But very less amount of these works have discussed the fact of 
externally controlled selective spin transfer through a nano-junction 
which is highly significant in designing controlled spintronics devices.} 
This essentially motivates us, and in the present work we intend to explore 
a possible route of getting externally controlled spin dependent transport.

We consider a simple two-terminal junction, where the bridging system is
a magnetic quantum ring. A finite modulation in site energy (described
by $\epsilon_i$, $i$ being the site index) is given in the form of
Aubry-Andr\'{e}-Harper (AAH) model~\cite{aubry,ds1,skm} i.e., $\epsilon_i =
w\cos(2i\pi\lambda+\phi_{\nu})$, where $w$ describe the width of the
site energy, and $\lambda$ is an irrational number which is fixed at
$(1 + \sqrt{5})/2$ (golden mean). The phase factor $\phi_{\nu}$ associated
with this expression plays an important role to regulate electron
transmission, more precisely, spin transmission. This $\phi_{\nu}$ {\em can 
be tuned externally}, which thus suggests a possible route of 
regulating spin transmission, without directly disturbing any other physical 
parameters. At the end of our theoretical analysis the 
feasibility of implementing such a model in laboratory is discussed.
Along with this, we propose another way of current regulation by
introducing the {\em proximity effect} of two non-magnetic source and drain
electrodes those are coupled to the neighboring sites of the ring (see
Fig.~\ref{fig1}). Due to close proximity an additional coupling is
established between the end atomic sites of the electrodes so that
electrons can directly tunnel between them~\cite{wg3,wg4} including their 
propagation through the magnetic quantum ring. This coupling which is of 
course tunable, plays a significant role in current regulation. From our 
numerical results we see that the present model exhibits a very high degree 
of spin polarization, some cases it almost reaches to $100\%$ and at the 
same time complete spin reversal can be achieved. Our results are valid 
for a wide range of parameter values, which demonstrates the robustness of
our proposition, and we strongly believe that both the two approaches can
be implemented experimentally.

\vskip 0.5cm
\noindent
{\bf Molecular Model and Theoretical Framework}
\vskip 0.3cm
\noindent
{\bf A. Model and Hamiltonian}
\vskip 0.25cm
\noindent
Let us begin with the nano-junction shown in Fig.~\ref{fig1} where a
$N$-site magnetic quantum ring is coupled to two perfect non-magnetic
semi-infinite metallic electrodes, namely, source and drain. Each site of
the ring is accompanied with a local magnetic moment with amplitude $h_i$
and its orientation is described by the polar angle $\theta_i$ and azimuthal
angel $\varphi_i$ in spherical polar coordinate system. At the same time, the
site energies get modified following the relation $w\cos(2i\pi\lambda +
\phi_{\nu})$ i.e., in the form of famous AAH model. Thus the bridging 
material is essentially a {\em correlated disordered system}, where the 
disorder is introduced only in site energy (viz, diagonal correlated 
disordered model). 

On the other hand, the two site-attached electrodes are perfect as well
as non-magnetic. Due to close proximity a direct coupling, described by
the parameter $t_C$, exists between the two end atomic sites of the
electrodes. This strength can be regulated either by changing the
separation between the electrodes or by rotating them~\cite{wg4}.

In order to write the Hamiltonian of the nano-junction we use Tight-Binding
(TB) framework which is extremely suitable for analyzing electron transport
particularly in the absence of electron-electron interaction. Within the
nearest-neighbor hopping approximation the Hamiltonian of the full system
looks like
\begin{equation}
\mathbf{H}=\mathbf{H_R} + \mathbf{H_S} + \mathbf{H_D} + \mathbf{H_T}
\label{eqn1}
\end{equation}
where different sub-Hamiltonians correspond to different parts as described
below.
The Hamiltonian of the magnetic quantum ring is written 
as~\cite{texp2,texp4,texp11}
\begin{eqnarray}
\mathbf{H_R} & = & \sum\limits_{i} \mbox{\boldmath $c_{i}^{\dagger}$}
\left(\mbox{\boldmath $\epsilon_{i}$} - \mbox{\boldmath $h_i$}.
\mbox{\boldmath $\sigma$}\right)\mbox{\boldmath $c_{i}$} + \sum\limits_{i}
\left(\mbox{\boldmath $c_{i+1}^{\dagger}$}\mbox{\boldmath $t_i$}
\mbox{\boldmath $c_{i}$} + h.c. \right)
\label{eqn2}
\end{eqnarray}
where, $\mbox{\boldmath $c_{i}$} = \left(\begin{array}{cc}
    c_{i\uparrow} \\ 
    c_{i\downarrow}
\end{array}\right)$, $\mbox{\boldmath $c_{i}^{\dagger}$}
= \left(\begin{array}{cc}
    c_{i\uparrow}^{\dagger} & c_{i\downarrow}^{\dagger}
\end{array}\right)$, 
$\mbox{\boldmath $t_{i}$}=\left(\begin{array}{cc}
    t & 0 \\ 
    0 & t
\end{array}\right)$,
$\mbox{\boldmath $\epsilon_{i}$}=\left(\begin{array}{cc}
    \epsilon_i & 0 \\ 
    0 & \epsilon_i
\end{array}\right)$,
$\mbox{\boldmath $h_i.\sigma$}=h_i\left(\begin{array}{cc}
    \cos\theta_i & \sin\theta_ie^{-j\varphi_i} \\ 
    \sin\theta_ie^{j\varphi_i} & -\cos\theta_i
\end{array}\right)$.
\vskip 0.2cm
\noindent
Here $t$ and $\epsilon_i$ correspond to the nearest-neighbor hopping 
(NNH) integral and site energy, respectively, in the ring. 
This site energy ($\epsilon_i$) is taken in the form of diagonal AAH 
model as discussed above. The term
$\mathbf{h_i}. \mbox{\boldmath $\sigma$}$ describes the interaction of
injected electron with the local magnetic moment placed at $i$-th site
having strength $h_i$.
$\mbox{\boldmath $\sigma$}\{=\mbox{\boldmath $\sigma_x$}, \mbox{\boldmath
$\sigma_y$}, \mbox{\boldmath $\sigma_z$}\}$ denotes the Pauli spin matrices
in $\mbox{\boldmath $\sigma_z$}$ diagonal representation.

The second and third sub-Hamiltonians in the right side of Eq.~\ref{eqn1} 
represent the source and drain electrodes, and they are expressed as
\begin{eqnarray}
\mathbf{H_S} & = & \sum\limits_{n \le -1}\mathbf{a_{n}
^{\dagger}}\mbox{\boldmath $\epsilon_{0}$}\mathbf{a_{n}} +
\sum\limits_{n \le -1}\left(\mathbf{a_{n}^{\dagger}}
\mathbf{t_0}\mathbf{a_{n-1}} + h.c. \right)
\label{eqn3}
\end{eqnarray}
and
\begin{eqnarray}
\mathbf{H_D} & = &\sum\limits_{n \ge 1}\mathbf{b_{n}^{\dagger}}
\mbox{\boldmath $\epsilon_{0}$}\mathbf{b_{n}} + \sum\limits_{n \ge 1}
\left(\mathbf{b_{n}^{\dagger}}\mathbf{t_0}\mathbf{b_{n+1}} + h.c.\right)
\label{eqn4}
\end{eqnarray}
where $a_n(b_n)$ and $a_n^{\dagger}(b_n^{\dagger})$ are the annihilation
and creation operators, respectively, for the source (drain) electrode.
The other symbols are
\vskip 0.2cm
\noindent
$\mbox{\boldmath $\epsilon_{0}$}=\left(\begin{array}{cc}
    \epsilon_0 & 0 \\ 
    0 & \epsilon_0
\end{array}\right)$,
$\mbox{\boldmath $t_{0}$}=\left(\begin{array}{cc}
    t_0 & 0 \\ 
    0 & t_0
\end{array}\right)$
\vskip 0.2cm
\noindent
where $\epsilon_0$ and $t_0$ are the site-energy and nearest-neighbor 
hopping integral in the electrodes, respectively.

Finally, $\mathbf{H_T}$, the tunneling Hamiltonian can be written as,
\begin{eqnarray}
\mathbf{H_T} & = & \left(\mathbf{c_{1}^{\dagger}}\mathbf{t_S}\mathbf{a_{-1}}
+\mathbf{c_{N}^{\dagger}}\mathbf{t_D}\mathbf{b_{1}} +
\mathbf{a_{-1}^{\dagger}}\mathbf{t_C}\mathbf{b_{1}} + h.c.\right)
\label{eqn5}
\end{eqnarray}
where, $\mbox{\boldmath $t_{K}$}=\left(\begin{array}{cc}
    t_K & 0 \\ 
    0 & t_K
\end{array}\right)\,$, $K=S,D,C$.
\vskip 0.2cm
\noindent
Here, $t_S$ and $t_D$ describe the couplings of the ring with source 
and drain, respectively and $t_C$ measures the direct coupling between 
the end atomic sites of the electrodes.

Below we discuss the theoretical prescription which includes the calculations
of spin dependent transmission probabilities, junction currents and spin
polarization.

\vskip 0.5cm
\noindent
{\bf B. Transmission Probability}
\vskip 0.25cm
\noindent
To calculate transmission probabilities we use wave-guide theory (which is
very simple to understand)~\cite{wg3,wg4,wg1,wg2}. {\em The theoretical 
prescription given below is an extension of earlier studies where spin 
degrees of freedom have not been taken into account.} Here we consider 
electron spin and the required steps are as follows.

Let us start with the station wave-function of the entire system (viz,
source-ring-drain)
\begin{equation}
|\psi\rangle =\left[\sum\limits_{n \le -1}\mbox{\boldmath $A_n$}
\mbox{\boldmath $a_{n,\sigma}^{\dagger}$} + \sum\limits_{n \ge 1}
\mbox{\boldmath $B_n$}\mbox{\boldmath $b_{n,\sigma}^{\dagger}$}
+ \sum\limits_{i=1}\mbox{\boldmath $C_i$}\mbox{\boldmath
$c_{i,\sigma}^{\dagger}$}\right]|0\rangle
\label{eqn6}
\end{equation}
where,
\vskip 0.2cm
\noindent
$\mbox{\boldmath $A_{n}$}=\left(\begin{array}{cc}
    A_{n,\uparrow} \\ 
    A_{n,\downarrow}
\end{array}\right)\,$, $\mbox{\boldmath $B_{n}$}=\left(\begin{array}{cc}
    B_{n,\uparrow} \\ 
    B_{n,\downarrow}
\end{array}\right)\,$, and $\mbox{\boldmath $C_{n}$}=\left(\begin{array}{cc}
    C_{n,\uparrow} \\ 
    C_{n,\downarrow}
\end{array}\right)\,$.
\vskip 0.2cm
\noindent
The coefficients $A_{n,\sigma}$, $B_{n,\sigma}$, and $C_{n,\sigma}$ correspond
to the amplitude for an electron having spin $\sigma$ ($\uparrow$ or
$\downarrow$) at the $n\,$th site of the source, drain, and $i\,$th site
of the ring, respectively. 

With this wave function we can write a set of coupled linear equations from
the time-independent Schr\"{o}dinger equation $\bf{H}|\psi\rangle$ $=$
$E\bf{I}|\psi\rangle$ ($\mathbf{I}$ being the ($2\times2$) identity matrix) 
as:
\begin{eqnarray}
\left(E\mathbf{I_2}-\mbox{\boldmath $\epsilon_{0,\sigma}$}\right)
\mathbf{A_n}&=&\mbox{\boldmath $t_{0,\sigma}$}\left(\mathbf{A_{n+1}} +
\mathbf{A_{n-1}}\right), n \leq -2, \nonumber \\
\left(E\mathbf{I_2}-\mbox{\boldmath $\epsilon_{0,\sigma}$}\right)
\mathbf{A_{-1}}&=&\mbox{\boldmath $t_{0,\sigma}$}\mathbf{A_{-2}} +
\mbox{\boldmath $t_{C,\sigma}$}\mathbf{B_1} + \mbox{\boldmath $t_{S,\sigma}$}
\mathbf{C_1},\nonumber \\
\left(E\mathbf{I_2}-\mbox{\boldmath $\epsilon_{0,\sigma}$}\right)
\mathbf{B_n} &=&\mbox{\boldmath $t_{0,\sigma}$}\left(\mathbf{B_{n+1}} +
\mathbf{B_{n-1}}\right),n\geq2,\nonumber \\
\left(E\mathbf{I_2}-\mbox{\boldmath $\epsilon_{0,\sigma}$}\right)\mathbf{B_1}
&=&\mbox{\boldmath $t_{0,\sigma}$}\mathbf{B_2} + \mbox{\boldmath
$t_{C,\sigma}$}\mathbf{A_{-1}} + \mbox{\boldmath $t_{D,\sigma}$}\mathbf{C_N},
\nonumber \\
\left(E\mathbf{I_2}-\mbox{\boldmath $\epsilon_{i,\sigma}$}\right)
\mathbf{C_i} & = &\mbox{\boldmath $t_{i,\sigma}$}\left(\mathbf{C_{i+1}} +
\mathbf{C_{i-1}}\right)
+ \mbox{\boldmath $t_{S,\sigma}$}\delta_{i,1}\mathbf{A_{-1}} \nonumber \\
 & & +~ \mbox{\boldmath $t_{D,\sigma}$}\delta_{i,N}\mathbf{B_1},~ 1\leq
i\leq N 
\label{eqn7}
\end{eqnarray}
\noindent
(i) \underline{\em{Up spin incidence from the source lead}}
\vskip 0.2cm
Assuming a plane wave incidence for up spin electrons with unit amplitude,
we can write the amplitudes as:
\vskip 0.2cm
\noindent
$\mbox{\boldmath $A_n$}=\left(\begin{array}{cc}
    e^{ik(n+1)a} + r_{\uparrow\uparrow}e^{-ik(n+1)a} \\
    r_{\uparrow\downarrow}e^{-ik(n+1)a} 
    \end{array}\right)$
and $\mbox{\boldmath $B_n$}=\left(\begin{array}{cc}
    t_{\uparrow\uparrow}e^{ikna} \\
    t_{\uparrow\downarrow}e^{ikna} 
    \end{array}\right)\,$,
\vskip 0.2cm
\noindent
where $a$ being the lattice spacing and $k$ is the wave vector associated with
the energy $E$. The other parameters are as follows:
\vskip 0.2cm
\noindent
$t_{\uparrow\uparrow}$ = Transmission amplitude of a up spin ($\uparrow$)
transmitted as up spin ($\uparrow$),\\
$t_{\uparrow\downarrow}$ = Transmission amplitude of a up spin ($\uparrow$)
transmitted as down spin ($\downarrow$).\\
$r_{\uparrow\uparrow}$ = Reflection amplitude of a up spin ($\uparrow$)
reflected as up spin ($\uparrow$),\\
$r_{\uparrow\downarrow}$ = Reflection amplitude of a up spin
($\uparrow$) reflected as down spin ($\downarrow$).
\vskip 0.2cm
\noindent
Using the expression of $\mathbf{A_n}$ and $\mathbf{B_n}$ we can now find
the reflection and transmission amplitudes by solving the set of coupled
equations (Eq.~\ref{eqn7}) for a particular energy associated with
each wave vector $k\,$. The we can define the pure spin transmission and spin
flip transmission probabilities as $T_{\uparrow\uparrow}=
|t_{\uparrow\uparrow}|^2$ and $T_{\uparrow\downarrow}=
|t_{\uparrow\downarrow}|^2$, respectively for the case of up spin incidence.
\vskip 0.2cm
\noindent
(ii) \underline{\em{Down spin incidence from the source lead}}
\vskip 0.2cm
For the case of down spin incidence the amplitudes $\mathbf{A_n}$ and
$\mathbf{B_n}$ look like:
\vskip 0.2cm
\noindent
$\mbox{\boldmath $A_n$}=\left(\begin{array}{cc}
    r_{\downarrow\uparrow}e^{-ik(n+1)a} \\
        e^{ik(n+1)a} + r_{\downarrow\downarrow}e^{-ik(n+1)a}
    \end{array}\right)$
and
$\mbox{\boldmath $B_n$}=\left(\begin{array}{cc}
    t_{\downarrow\uparrow}e^{ikna} \\
    t_{\downarrow\downarrow}e^{ikna} 
    \end{array}\right)\,$,
\vskip 0.2cm
\noindent
where the meaning of different factors are as follows:
\vskip 0.2cm
\noindent
$t_{\downarrow\uparrow}$ = Transmission amplitude for down spin ($\downarrow$)
transmitted as up spin ($\uparrow$),\\
$t_{\downarrow\downarrow}$ = Transmission amplitude for down spin
($\downarrow$) transmitted as down spin ($\downarrow$).\\
$r_{\downarrow\uparrow}$ = Reflection amplitude for down spin ($\downarrow$)
reflected as up spin ($\uparrow$),\\
$r_{\downarrow\downarrow}$ = Reflection amplitude for down spin
($\downarrow$) reflected as down spin ($\downarrow$).
\vskip 0.2cm
\noindent
Using the same prescription as stated for the case of up spin incidence,
here also we can calculate all coefficients by solving the equations given
in Eq.~\ref{eqn7}, and eventually, find the transmission probabilities as
$T_{\downarrow\downarrow}=|t_{\downarrow\downarrow}|^2$ and
$T_{\downarrow\uparrow}=|t_{\downarrow\uparrow}|^2$.

Finally we can write the total transmission probability for spin up as
$T_{\uparrow} = T_{\uparrow\uparrow} + T_{\downarrow\uparrow}$ and
for spin down as $T_{\downarrow} = T_{\uparrow\downarrow} + 
T_{\downarrow\downarrow}$.

\vskip 0.5cm
\noindent
{\bf C. Junction Current}
\vskip 0.25cm
\noindent
Once the transmission function is determined, the net junction current
for a particular bias voltage $V$ at absolute zero temperature, can be
evaluated from the relation~\cite{datta1}
\begin{equation}
I_{\sigma}(V) = \frac{e}{\pi \hbar} \int\limits_{E_F-\frac{eV}{2}}^{E_F+
\frac{eV}{2}}T_{\sigma}(E) \, dE
\label{eqn8}
\end{equation}
where $E_F$ is the equilibrium Fermi energy.

\vskip 0.5cm
\noindent
{\bf D. Spin Polarization}
\vskip 0.25cm
\noindent
Finally, we define spin polarization coefficient as~\cite{pola}
\begin{equation}
P = \frac{I_{\uparrow} - I_{\downarrow}}{I_{\uparrow} + I_{\downarrow}}
\label{eqn9}
\end{equation}
$P=+1(-1)$ corresponds to the only up (down) spin 
propagation, and thus, under this situation the degree of up (down) 
spin polarization becomes $100\%$. $P=0$ represents no spin polarization.

\vskip 0.5cm
\noindent
{\bf Numerical Results and Discussion}
\vskip 0.2cm
\noindent
Following the above theoretical prescription now we present our numerical
results. The physical parameters those are kept constant throughout the
computation are as follows. In the source and drain 
electrodes, the site energy $\epsilon_0$ and nearest-neighbor hopping 
integral $t_0$ are fixed at $0$ and $3\,$eV, respectively, whereas in 
the bridging conductor (i.e., the ring) we set $t=1\,$eV and choose 
$\epsilon_i$ following the relation $\epsilon_i=w \cos\left(2 i \pi \lambda
+\phi_{\nu}\right)$ considering $w=1\,$eV. In the ring conductor we consider
the strength of magnetic moment $h_i=1\,$eV and the azimuthal angle
$\varphi_i=0$ for all $i$ and also, unless otherwise specified, $\theta_i=0$ 
for all $i$. The other two parameters $t_S$ and $t_D$ are fixed at $1\,$eV. 
The values of $t_C$ and phase factor $\phi_{\nu}$ are placed in appropriate 
figures, as they are not constant. All the calculations presented below 
are computed at absolute zero temperature setting equilibrium Fermi energy 
$E_F =0$.

Before addressing the central issues i.e., regulations of spin polarization
\begin{figure}[ht]
{\centering \resizebox*{8cm}{6cm}{\includegraphics{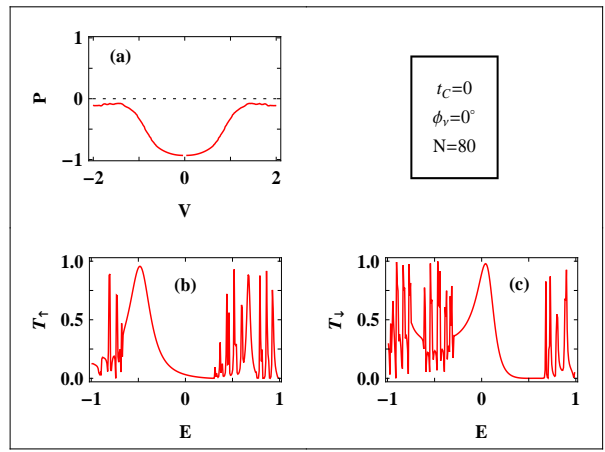}}\par}
\caption{(Color online). Voltage dependent spin polarization coefficient 
$P$ along with spin dependent transmission probabilities $T_{\uparrow}$ and
$T_{\downarrow}$ as a function of injecting electron energy $E$ for a 
$80$-site ring at some typical values of $\phi_{\nu}$ and $t_C$. At zero 
bias ($V=0$) there is no current across the junction, and thus, we cannot
take the ratio of the currents following Eq.~\ref{eqn9} as it is undefined. 
Therefore, we ignore this point in the $P$-$V$ curve.}
\label{f2e1}
\end{figure}
\begin{figure}[ht]
{\centering \resizebox*{8cm}{6cm}{\includegraphics{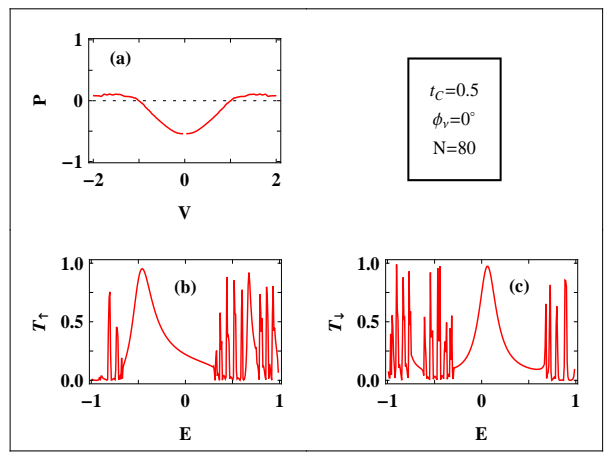}}\par}
\caption{(Color online). Same as Fig.~\ref{f2e1}, with $t_C=0.5\,$eV.}
\label{f2e2}
\end{figure}
\begin{figure}[ht]
{\centering \resizebox*{8cm}{6cm}{\includegraphics{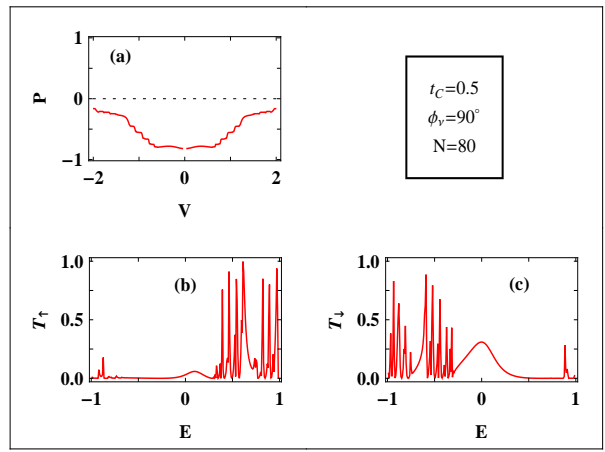}}\par}
\caption{(Color online). Same as Fig.~\ref{f2e1}, with $t_C=0.5\,$eV
and $\phi_{\nu}=\pi/2$.}
\label{f2e3}
\end{figure}
as well as spin inversion with the help of external phase $\phi_{\nu}$
and direct coupling parameter $t_C$, let us start by analyzing spin
polarization coefficient for some typical values of $t_C$ and $\phi_{\nu}$.
The results are presented in Figs.~\ref{f2e1}-\ref{f2e3}, where the variation
of spin polarization $P$ is given as a function of bias voltage $V$ along
with up and down spin transmission probabilities considering the ring size 
$N=80$.

For $t_C=0$ and $\phi_{\nu}=0\,$, the spin polarization coefficient $P$
almost reaches to a maximum for low bias region ($P=+1$ or $P=-1$ represents
a maximum spin polarization associated with the complete suppression of
down or up spin propagation through the junction), and it ($P$) gradually
decreases with increasing bias voltage and eventually drops almost
to zero for higher voltages (Fig.~\ref{f2e1}(a)). This behavior can be
justified from the transmission spectra given in Figs.~\ref{f2e1}(b) and
(c). For narrow energy window across $E=0\,$, transmission probability of
up spin electrons is almost zero whereas finite transmission of other spin
electrons is obtained which results $P\sim-1$ over a narrow voltage region
associated with the energy window.
But when we consider wide energy region, associated with the bias voltage,
both up and down spin channels contribute in electronic transmissions
yielding lesser spin polarization.

The spin polarization, more precisely spin selective transmission, essentially
depends on the separation between up and down spin channels. For such a
system, where atomic sites of the bridging conductor are magnetic, spin
flip interaction term is responsible for it. As hopping integral is
fixed (same for both up and down spin electrons), the separation
between the up and down spin channels is controlled by the term
$\left(\epsilon_{i} - h_i.\sigma\right)\,$, out of which $\epsilon_{i}$ 
again contains a tunable factor $\phi_{\nu}$ and its precise role can be 
understood from the forthcoming analysis.

Apart from this factor (i.e., $\epsilon_{i}-h_i.\sigma\,$), quantum 
interference 
has significant role on spin selective transmission. To reveal this
fact let us focus on the results placed in Fig.~\ref{f2e2}(a), where we set
a finite $t_C\,$, keeping all other parameters unchanged as taken in
Fig.~\ref{f2e1}(a). Introduction of $t_C$ means there is an addition of 
\begin{figure}[ht]
{\centering \resizebox*{7cm}{9.5cm}{\includegraphics{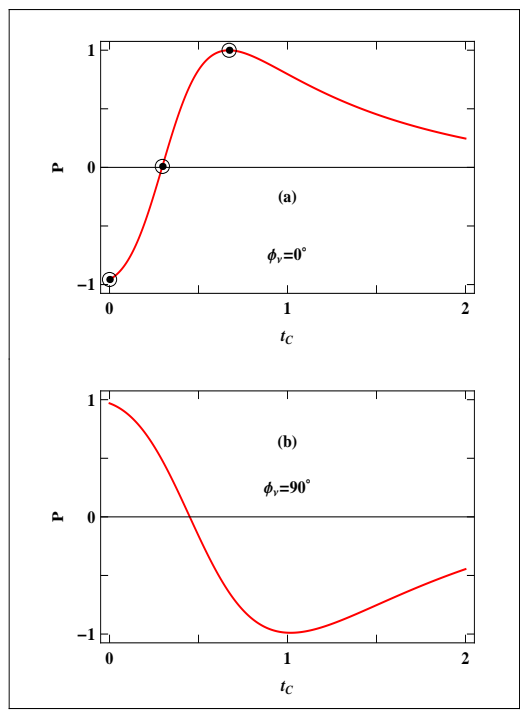}}\par}
\caption{(Color online). Spin polarization coefficient $P$ as a function 
of $t_C$ for a $120$-site ring considering two different values of 
$\phi_{\nu}$. Here we set $V=0.25\,$V.}
\label{f3}
\end{figure}
a new path along with two conducting paths (namely longer and shorter paths 
in the magnetic quantum ring). Thus, these three paths are responsible 
for electronic transmission and we get the combined effect in the drain 
electrode. In presence of $t_C$ the degree of spin polarization gets 
reduced, compared to the previous case (viz, Fig~\ref{f2e1}(a)), which 
is clearly noticeable in the low bias region (Fig.~\ref{f2e2}(a)). This 
reduction of spin polarization is expected because of the inclusion of 
new path which allows in certain percentage to pass up and down spin 
electrons, avoiding the magnetic ring. This is reflected in the 
transmission-energy spectra where we get finite transmission probabilities 
for both up and down spin electrons. So for a particular voltage window 
both of them are contributing, and depending on the contributing electrons 
we get a net polarization (which of course is less than $100\,\%$). 
For large enough $t_C$, one can expect much lesser spin polarization for 
any bias window as in that case electrons directly pass through this new 
path, without encountering any spin dependent interaction in the magnetic 
ring.

Under this situation if we incorporate the phase factor $\phi_{\nu}$ then
transmission spectra for both up and down spin electrons get modified,
(Figs.\ref{f2e3}(b) and (c)), and accordingly, spin polarization changes
(Fig.~\ref{f2e3}(a)). Around $80\,\%$ spin polarization is achieved for a 
wide bias window, though eventually it decreases with higher voltages like 
the other two cases (viz, Figs.~\ref{f2e1} and \ref{f2e2}).

From the results analyzed so far (i.e., Figs.~\ref{f2e1}-\ref{f2e3}), we 
see that in the low bias region down spin electrons dominate suppressing
\begin{figure}[ht]
{\centering \resizebox*{7cm}{13cm}{\includegraphics{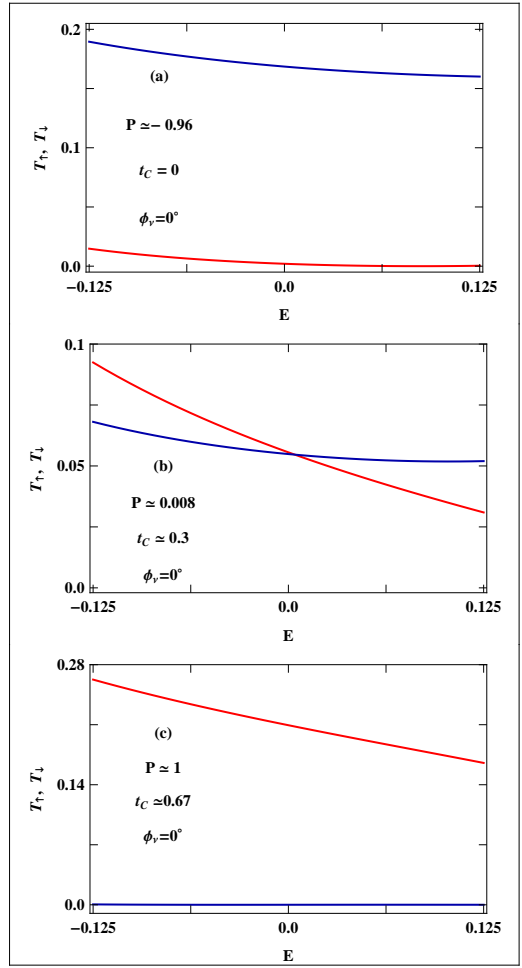}}\par}
\caption{(Color online). Energy dependence of $T_{\uparrow}$ (red curve) 
and $T_{\downarrow}$ (blue curve) at three different values $t_C$ those
are represented by encircled dots in Fig.~\ref{f3}(a). The other physical
parameters are same as taken in Fig.~\ref{f3}.}
\label{f4}
\end{figure}
the other spin electrons. An exactly opposite behavior might be observed for
other set of parameter values depending on the channel separation, which 
in principle, is regulated by several factors for the present model.

\vskip 0.2cm
\noindent
{\bf $\bullet$ Regulation of spin polarization by $t_C$}
\vskip 0.2cm
\noindent
Now we discuss the explicit dependence of spin polarization $P$ on the 
coupling parameter $t_C$. The results are presented in Fig.~\ref{f3} for
a $120$-site ring considering two different values of $\phi_{\nu}$.
Two observations are noteworthy. First, by regulating the external tunneling
coupling $t_C$, $P$ can be changed widely from $+1$ to $-1$ and vice versa.
Second, a phase reversal of spin polarization takes place with the help of
AAH phase $\phi_{\nu}$. When $\phi_{\nu}=0$, $P$ varies from $-1$ to $+1$,
while for the other case ($\phi_{\nu}=\pi/2$), it ($P$) runs from $+1$ to 
$-1$, and for large $t_C$ decreasing spin polarization is observed in these 
two cases.

To implement this wide variation of $P$, we choose three distinct points 
from $P$-$t_C$ curve of Fig.~\ref{f3}(a), represented by encircled dots, 
\begin{figure}[ht]
{\centering \resizebox*{7cm}{9.5cm}{\includegraphics{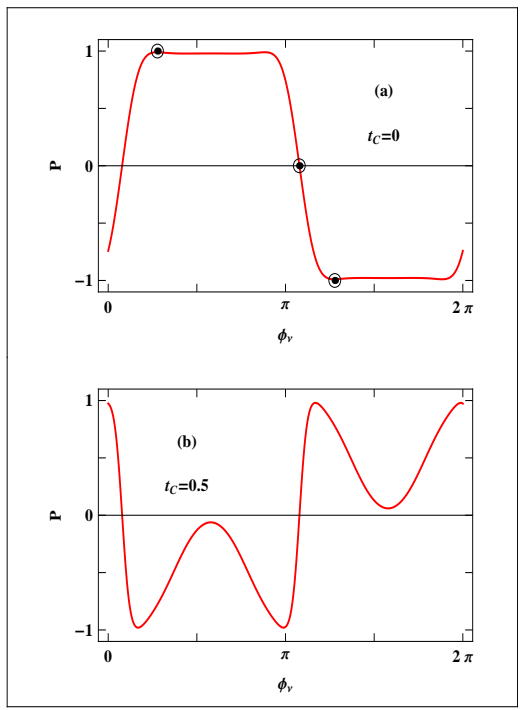}}\par}
\caption{(Color online). $P$-$\phi_{\nu}$ characteristics at two typical
values of $t_C$ for a $100$-site ring considering $V=0.25\,$V.}
\label{f5}
\end{figure}
and present the characteristics of up and down spin transmission 
probabilities for these $t_C$ in Fig.~\ref{f4}. The results are shown for a
specific energy window (-$0.125 \le E \le 0.125$) associated with the 
voltage $V=0.25\,$V. When $\phi_{\nu}=0$ and $t_C=0$, up spin transmission
probability is almost zero (red line of Fig.~\ref{f4}(a)), while finite
transmission probability is obtained for down spin electrons (blue line
of Fig.~\ref{f4}(a)) which results $P \sim -1$. The scenario gets reversed
at $t_C\simeq0.67$, shown in Fig.~\ref{f4}(c), where only up spin electrons 
transmit through the junction providing $P=+1$. At $t_C\simeq0.3$, finite 
transmission probabilities are obtained for both up and down spin electrons,
and $I_{\uparrow}$ is very close to $I_{\downarrow}$ which gives vanishing
spin polarization (Fig.~\ref{f4}(b)). Similar kind of analysis is also used
for analyzing the behavior of spin polarization in the system with 
$\phi_{\nu}=\pi/2$.

In addition to these features it is also observed that for both zero and
non-zero values of AAH phase, $P$ gradually decreases with increasing $t_C$
as two opposite spin electrons are allowed to pass more easily from the 
source to drain electrode without encountering magnetic region. Thus, from
the results presented in Fig.~\ref{f3}, it can be emphasized that 
{\em controlling $t_C$ externally, the spin polarization can be varied in 
a wide range ($+1$ to $-1$ and vice versa) through this nano-junction, 
without changing any other physical parameters.} This is indeed an 
interesting observation and we believe that it can be verified through 
an experimental setup.

\vskip 0.2cm
\noindent
{\bf $\bullet$ Regulation of spin polarization by $\phi_{\nu}$}
\vskip 0.2cm
\noindent
To establish the specific dependence of $P$ on phase factor $\phi_{\nu}$,
in Fig.~\ref{f5} we present the results for a $100$-site ring considering 
two typical values of $t_C$. 
\begin{figure}[ht]
{\centering \resizebox*{7cm}{12cm}{\includegraphics{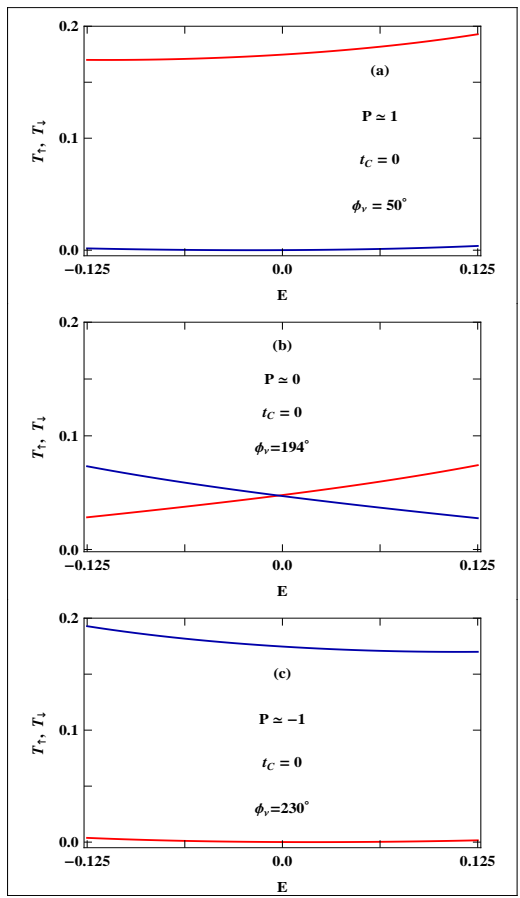}}\par}
\caption{(Color online). Energy dependence of $T_{\uparrow}$ (red curve) 
and $T_{\downarrow}$ (blue curve) at three different values $\phi_{\nu}$ 
those are represented by encircled dots in Fig.~\ref{f5}(a). The other 
physical parameters are kept constant as taken in Fig.~\ref{f5}.}
\label{f6}
\end{figure}
\begin{figure}[ht]
{\centering \resizebox*{7.5cm}{6cm}{\includegraphics{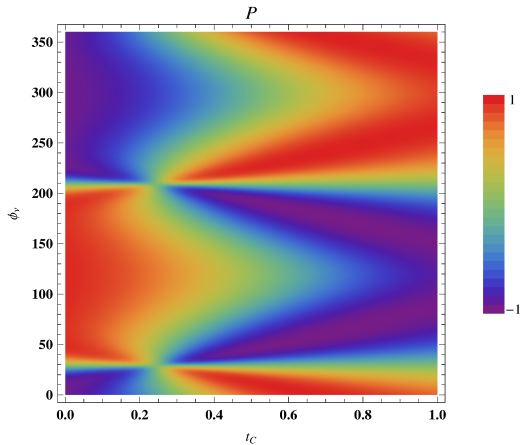}}\par}
\caption{(Color online). Simultaneous variation of $P$ with $t_C$ and
$\phi_{\nu}$ for a $60$-site ring at $V=0.25\,$V.}
\label{f7}
\end{figure}
Quite interestingly we see that, like Fig.~\ref{f3}, here also the spin
polarization coefficient exhibits a wide range of variation (cent percent
up spin polarization to cent percent down spin polarization and vice versa)
upon the change of $\phi_{\nu}$ for a fixed $t_C$. The role of $t_C$ on
phase reversal is also clear from the spectra given in Figs.~\ref{f5}(a)
and (b). This interesting pattern can be visualized from the transmission
spectra placed in Fig.~\ref{f6}, where we present the variations of up and
down spin transmission probabilities in a particular energy window 
associated with the voltage bias, selectively choosing three arbitrary
points from the $P$-$\phi_{\nu}$ curve of Fig.~\ref{f5}(a), represented 
by encircled dots, where $P$ becomes $\sim +1$, $0$ and $-1$, respectively.
For a particular phase a situation may arise where only up spin electrons 
transmit resulting $P=+1$, and the other situation can also happen for 
another phase value where only down spin electrons propagate yielding $P=-1$.
The third possibility is that for a specific $\phi_{\nu}$ both electrons 
can contribute equally in a typical voltage window providing vanishing 
transmission probability. All these possible cases are visualized clearly 
from Fig.~\ref{f5}. {\em Since 
this phase factor $\phi_{\nu}$ is tuned externally, we can suggest that 
the present model can be utilized as a phase controlled device for getting 
selective spin transmission through a nano-junction.}

Like the case of controlling spin polarization by 
introducing $t_C$, one may think whether there is any possibility to expect 
the {\em wide variation of spin polarization} as a function of phase factor
$\phi_{\nu}$ without doing any numerical calculations or not. The answer 
is of course yes, since it depends on which spin channel (up or down) is 
dominating the other for a specific energy window associated with bias 
voltage $V$. The widths of up and down spin bands of the magnetic quantum 
ring essentially depends on the factors $\epsilon_i$, $h$ and NNH integral 
$t$. Based on these parameter values we get an overlap between the two spin
bands over a finite energy window, while no overlap is obtained for other 
energy regions. This overlapping region, on the other hand, can be 
controlled by tuning the phase factor $\phi_{\nu}$ as it eventually 
regulates the site energy $\epsilon_i$ through a cosine modulation term.
Thus, for a fixed Fermi energy, when overlap region comes within a voltage
window for a specific $\phi_{\nu}$, vanishingly small spin polarization is
observed, whereas keeping all other parameters unchanged we can shift the 
overlap region from the voltage window by tuning $\phi_{\nu}$ and in that
case high degree of up (down) spin polarization is obtained depending on the
specific channel. This is exactly what we see in Fig.~\ref{f5}.

It is to
be noted that when all site energies ($\epsilon_i$'s) are same i.e., the 
system becomes an ordered magnetic ring, the eigenenergies of up and down 
spin bands can be evaluated analytically so that their overlap can easily 
be estimated. While, for correlated site energies (like our present model) 
analytical solution is no longer available. Though we can intuitively 
estimate the wide variation of spin polarization with phase $\phi_{\nu}$ 
without doing numerical calculations, complete transmission-energy spectrum 
only reveals the precise determination of spin polarization at different 
phases.

\vskip 0.2cm
\noindent
{\bf $\bullet$ Simultaneous variation of $P$ by $t_C$ and $\phi_{\nu}$}
\vskip 0.2cm
\noindent
From the above analysis (Figs.~\ref{f3}-\ref{f6}) naturally the question
appears how the spin polarization gets modified with the simultaneous 
\begin{figure}[ht]
{\centering \resizebox*{7cm}{6cm}{\includegraphics{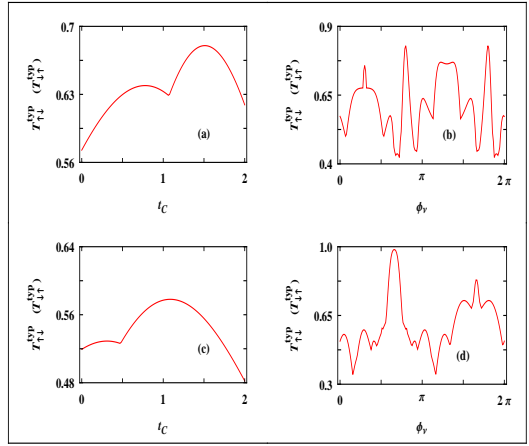}}\par}
\caption{(Color online). Spin flip transmission probabilities 
$T_{\uparrow\downarrow}^{\mbox{\tiny typ}}$
($T_{\downarrow\uparrow}^{\mbox{\tiny typ}}$) as a function of $t_C$ 
($\phi_{\nu}$) for two different ring sizes, where the upper and lower rows
correspond to $N=60$ and $40$, respectively. Here we choose $\theta_i=\pi/2$
$\forall$ $i$. The typical value of spin-flip transmission probability is 
determined by taking the maximum value of $T_{\sigma\sigma^{\prime}}$ from 
the $T_{\sigma\sigma^{\prime}}$-$E$ curve considering the variation of $E$
within the energy window $-4\leq E \leq 4$.}
\label{f8}
\end{figure}
\begin{figure}[ht]
{\centering \resizebox*{7.5cm}{6.0cm}{\includegraphics{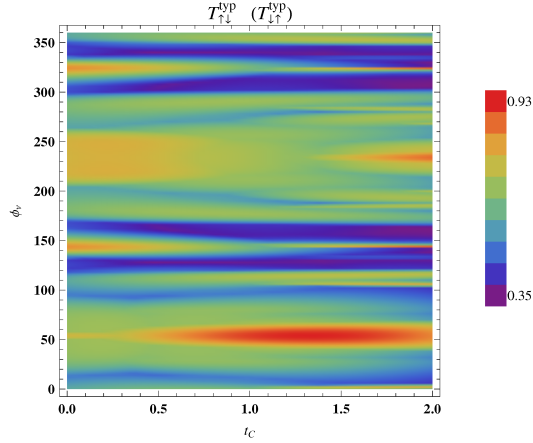}}\par}
\caption{(Color online). Simultaneous variation of 
$T_{\uparrow\downarrow}^{\mbox{\tiny typ}}$
($T_{\downarrow\uparrow}^{\mbox{\tiny typ}}$) with $t_C$ and $\phi_{\nu}$
for a $60$-site ring considering $\theta_i=\pi/2$ $\forall$ $i$. 
$T_{\sigma\sigma^{\prime}}$ is determined in the same way like Fig.~\ref{f8}.}
\label{f9}
\end{figure}
variation of both $t_C$ and $\phi_{\nu}$. The answer is given in 
Fig.~\ref{f7} where we present the dependence of $P$ as functions of
$t_C$ and $\phi_{\nu}$ considering a $60$-site ring at $0.25\,$Volts.
This is a clear picture to visualize the combined role of these two 
externally controlling parameters. For lower $t_C$, $P$ becomes $\sim +1$
or $\sim -1$ for a wide range of $\phi_{\nu}$ providing a broad zone
of identical color (red or pink), while the width of these zones becomes 
narrow down as we move towards higher $t_C$. This diagram suggests that
the physical pictures are valid over a large range of parameter values,
rather than a specific $t_C$ and $\phi_{\nu}$, which claims the robustness 
of our observation.

\vskip 0.2cm
\noindent
{\bf $\bullet$ Spin Inversion}
\vskip 0.2cm
\noindent
Finally, we concentrate on spin-flip scattering through this nano-junction.
To get spin-flip transmission we have to set a non-zero value of $\theta_i$,
as $\theta=0^{\circ}$ (we can call $\theta_i=\theta$ $\forall$ $i$, for
simplification) does not involve the factors $\sigma_+$ and $\sigma_-$
in the spin-flip term $\vec{h}_i.\vec{\sigma}$ (Eq.~\ref{eqn2})~\cite{mdflip}
which are responsible for spin flipping.

In Fig.~\ref{f8} we present the spin-flip transmission probabilities 
$T_{\uparrow\downarrow}^{\mbox{\tiny typ}}$ 
($T_{\downarrow\uparrow}^{\mbox{\tiny typ}}$) for two different ring 
sizes considering $\theta=\pi/2$, where the upper and lower rows correspond
to $N=60$ and $40$, respectively. The typical value of spin-flip transmission
probability is determined by taking the maximum value of 
$T_{\sigma\sigma^{\prime}}$ from the $T_{\sigma\sigma^{\prime}}$-$E$ curve
considering the variation of $E$ within the energy window $-4\leq E \leq 4$.
From the spectra it is observed that for a finite (small) window of AAH
phase a complete spin reversal takes place (see Fig.~\ref{f8}(d)), while
in other cases though full spin inversion is not available but the degree
of spin inversion is sufficiently high at some particular $t_C$ and 
$\phi_{\nu}$ windows (Fig.~\ref{f8}). It indicates that by controlling the 
physical parameters a possibility may arise to achieve complete spin 
inversion through this nano-junction.

To make it more clear in Fig.~\ref{f9} we present  
$T_{\uparrow\downarrow}^{\mbox{\tiny typ}}$ 
($T_{\downarrow\uparrow}^{\mbox{\tiny typ}}$) as functions of both 
$\phi_{\nu}$ and $t_C$ considering $N=60$ and $\theta=\pi/2$. Almost 
$95\,\%$ spin inversion takes place for a reasonable window of the parameter
values which definitely suggests an experimental verification as the results
are not so sensitive with fine tuning of these parameters. In addition,
we would like to state that though the results presented in Fig.~\ref{f9}
are computed for a specific value of $\theta$, almost similar kind of 
physical picture (viz, large degree of spin polarization for wide window
of parameter values) is also obtained for other values of $\theta$.
Therefore, we do not repeat the same thing considering different values 
of this parameter ($\theta$).

\vskip 0.2cm
\noindent
{\bf $\bullet$ Experimental Perspective}
\vskip 0.2cm
\noindent
In order to substantiate the proposed scheme of tuning 
spin polarization via controlling the phase factor $\phi_{\nu}$ in 
laboratory we have to think about the possible realization of an 
experimental setup.
\begin{figure}[ht]
{\centering \resizebox*{5cm}{4cm}{\includegraphics{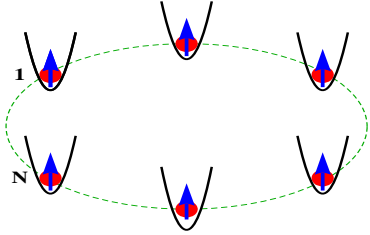}}\par}
\caption{(Color online). Schematic view of a ring-shaped 
geometry where 
trapping potentials are formed by two laser beams. In each such potentials
(described by black line) a magnetic atom is trapped to form a magnetic 
quantum ring with modulation in site energy. The source and drain electrodes
will be connected at the sites $1$ and $N$, respectively.}
\label{exp}
\end{figure}
Our essential goal is to develop a $1$D magnetic quantum 
ring where site energies are modulated in the form of standard AAH model
i.e., in one hand the site energies are quasiperiodic and in the other hand
this deterministic energy profile can be regulated externally (which is 
$\phi_{\nu}$ in our model). Several experimental proposals have been made 
along this direction to construct such a ring-like geometry, in fact 
different other geometrical shapes can also be 
designed~\cite{oexp1,oexp2,oexp3,oexp4,oexp5}. Two counter 
propagating laser beams having wave vectors $k_1$ and $k_2$ are used for 
generating such a quasiperiodic potential, where the incommensuration
parameter is defined by the factor $k_1/k_2$. Once the profile is formed 
by optical means then magnetic atoms are trapped in the dip regions as 
shown in Fig.~\ref{exp}. Tuning any one the two laser beams the profile 
can be regulated which practically describes the change of phase factor 
$\phi_{\nu}$ externally. Thus, a magnetic quantum ring with finite 
modulation in site energies can be formed through which spin-dependent 
transport can be tested. The details of experimental realization are 
available in Refs.~\cite{oexp1,oexp2,oexp3,oexp4,oexp5}. Before we end,
we would like to point out that with the help of interfering laser beams
different kinds of aperiodic lattices (our model is one such case) can 
be formed, but it is very hard to design a setup to map a random 
disordered model since in this case site energies are no longer 
correlated.

The other scheme of spin current regulation by means of
tuning $t_C$ can easily be implemented in a laboratory setup. One can do
it either by changing the separation between the source and drain electrodes 
or by rotating them~\cite{wg4}.

\vskip 0.5cm
\noindent
{\bf Summary}
\vskip 0.2cm
\noindent
To conclude, in the present work {\em two new mechanisms} have been pointed 
out for the regulation of spin polarization as well as spin inversion through
a magnetic nano-junction. A complete sign reversal of spin polarization (i.e.,
$P=+1$ to $P=-1$ and vice versa) takes place by changing any one of the 
two controlling parameters (viz, $\phi_{\nu}$ and $t_C$). The tunneling 
coupling $t_C$ between the electrodes can be regulated {\em externally} 
by some mechanical ways, and the other physical parameter i.e., AAH phase 
$\phi_{\nu}$ can also be tuned {\em externally}. Our results are valid for 
a wide range of parameter values, and thus, definitely an experimental 
verification can be made along this line. Focusing in that direction, 
finally we have discussed briefly how the proposed model can be realized 
in laboratory.

We have given a detailed theoretical description for the calculation of 
spin dependent transmission probabilities based on quantum wave-guide 
theory which might be helpful for investigating spin dependent transport 
through any such magnetic system. The scattering theory presented here
is the extension of earlier studies where spin degrees of freedom have been
ignored. So, in that context our theoretical prescription based on 
wave-guide theory involving electron spin is quite new, to the best of 
our knowledge.

In our forthcoming work we will analyze the behavior of spin polarization 
in such a nano-junction where two different phases, namely $\phi_{\nu}$ and
$\phi_{\lambda}$, are introduced in site potentials and hopping integrals,
respectively, along with the external tunneling coupling $t_C$. Both these
phases ($\phi_{\nu}$ and $\phi_{\lambda}$) can be regulated simultaneously
and independently through an experimental setup, and we strongly believe 
that some interesting features will be obtained that can be utilized 
in designing spin based quantum devices.

\vskip 0.2cm
\noindent
{\bf $\bullet$ Some Additional Points}
\vskip 0.2cm
\noindent
Here we would like to discuss some additional points for 
the sake of completeness and the benefit of interested researchers.
\vskip 0.2cm
\noindent
A. In our model we have considered identical strength of all
magnetic moments (i.e., $h_i=h$ (say) for all $i$). One can in principle
consider different $h_i$ which means different magnetic sites in the ring.
The main reason of not considering different $h_i$ is that here we intend 
to focus on the interplay between correlated diagonal disorder (that can 
be designed experimentally) and the external coupling (shunting path) term 
$t_C$. So there are two factors (i) phase in site potential and (ii) $t_C$, 
that can be used to regulate spin transmission through the conducting
junction. Introduction of different $h_i$ does not provide any new physical
signature. 
Only the height of the transmission peaks get reduced without changing the 
polarization characteristics. The same argument also goes to select the 
other two parameter values ($\theta_i$ and $\varphi_i$).
\vskip 0.2cm
\noindent
B. In describing the Hamiltonian of the magnetic quantum 
ring (Eq.~\ref{eqn2}) we have ignored exchange interaction term between
local magnetic moments. So one may ask why we have not considered the 
exchange term. The reason is that at low temperature this interaction 
term has very minor impact and does not make any qualitative difference. 
And the other important point is that since thermal broadening of energy 
levels is too weak compared to the energy level broadening caused by 
ring-to-electrode coupling, even moderate temperature are expected to 
have a very little impact on our qualitative predictions~\cite{datta1}. 
Therefore only zero temperature has been considered here. Naturally at 
zero temperature we can ignore this interaction term.
\vskip 0.2cm
\noindent
C. It is well known that Rashba SO coupling is responsible 
for spin-flip scattering. So the question naturally comes can we expect 
similar kind of characteristic features, as discussed above, if we replace 
the magnetic quantum ring by a Rashba ring. The answer is of course no. The 
first thing is that in a two-terminal system only SO coupling is not 
responsible for producing polarized spin currents. We have to apply a
magnetic field to break the Kramer's degeneracy, and confining of a magnetic
field in a small sized ring is always a difficult task. This part has already 
been discussed in the introduction.

The other point is that it is
very hard to design a Rashba ring considering such a deterministic
disordered potential in experiment, whereas magnetic atomic sites can 
easily be trapped optically. The Rashba term appears because of the 
asymmetry in the confining potential. So the mechanism is completely 
different and we do not know whether it is at all possible to design 
a Rashba ring by constructing a potential profile with the help of two 
interfering laser beams. May be a theoretical analysis can be done using 
a two-terminal Rashba ring in presence of magnetic field or considering 
a three-terminal Rashba ring (where magnetic field is no longer required 
to get spin polarization in outgoing leads) by this same prescription, 
but question may arise how to design such a model experimentally.

\vskip 0.2cm
\noindent
D. Throughout the numerical analysis we set a specific 
parameter values of $w$, $t_S$, $t_D$ and $t$. Naturally the question
may arise how the results get modified if we choose other set of parameter
values, for example, if we increase or decrease $w$, $t_S$, $t_D$ compared 
to $t$.

First consider the effect of $w$ and (say) we are 
increasing $w$. It ($w$) measures the correlated disorder strength. So 
keeping all other parameters fixed if we increase $w$ then disorder 
strength will be increased which means electronic states will be less 
conducting, as expected in correlated disordered systems. Accordingly 
peak heights 
in transmission spectra get reduced. So eventually for large enough 
disorder strength ($w>>t$) all states of the ring will be almost 
localized. Under this situation electrons will not enter into the 
ring geometry. But due to the additional shunting path, which is 
incorporated by considering a coupling between two electrodes, electron 
can easily hop from source to drain, avoiding the localized regime
i.e., the ring geometry. As the electrons are not entering into the ring
they will not experience any spin-dependent scattering and hence for this
large enough $w$ we will not get any spin polarization.

We can also think the above situation in other way. 
Suppose we fix $w$ which is not so large to localize electrons. Under 
this situation if we increase the coupling term $t_C$ then electrons will 
try to pass directly from source to drain, ignoring the ring geometry. 
In that case also we get decreasing spin polarization. Since disorder effect
is well known we do not want to repeat this, whereas we present our results
by changing $t_C$ which on the other hand can be realized in experiments 
quite easily.

Now we discuss the case where $w$ gets decreased. In this 
case electron will try to move through the ring, and there are two possible 
paths in the ring. So in total three possible paths: two arms in the ring 
(say upper and lower arms) and the third one is the shunting path. Thus
combined interference effect will be there which again analogous to the 
change of $t_C$ for a fixed $w$. Because of this, we have elaborately 
described the effect of coupling $t_C$.

Finally, we focus on the ring-to-electrode coupling effect 
i.e., how the results get affected by changing $t_S$ and $t_D$ with respect 
to $t$. This coupling effect has already been studied in a series of papers 
by us and other few authors too. Therefore, we do not want to repeat this 
behavior once again, and one can easily follow this effect from the 
Refs.~\cite{texp2,wg3,ws1,ws2,ws3}.

\vskip 0.35cm
\noindent
{\bf Acknowledgement}
\vskip 0.15cm
\noindent
MP is thankful to University Grants Commission (UGC), India
(F. 2-10/2012(SA-I)) for research fellowship. SKM is thankful to 
Prof. Abraham Nitzan for meaningful discussions. MP and SKM would like 
to thank the reviewers for their valuable comments and suggestions 
to improve the quality of the work.

\vskip 0.35cm
\noindent
{\bf Author Contributions}
\vskip 0.15cm
\noindent
S.K.M. conceived the project. M.P. performed numerical calculations.
M.P. and S.K.M. analyzed the data. S.K.M. supervised the theoretical
calculations. M.P. and S.K.M. co-wrote the paper.

\vskip 0.35cm
\noindent
{\bf Additional Information}
\vskip 0.15cm
\noindent
Correspondence should be addressed to S.K.M.

\vskip 0.35cm
\noindent
{\bf $^*$Correspondence to:}
santanu.maiti@isical.ac.in

\vskip 0.35cm
\noindent
{\bf Competing financial interests}
\vskip 0.15cm
\noindent
The authors declare no competing financial interests.

\end{document}